\documentclass[fleqn,usenatbib,useAMS]{mnras}

\usepackage{newtxtext,newtxmath}

\usepackage[T1]{fontenc}

\DeclareRobustCommand{\VAN}[3]{#2}
\let\VANthebibliography\thebibliography
\def\thebibliography{\DeclareRobustCommand{\VAN}[3]{##3}\VANthebibliography}

\usepackage{graphicx}	
\usepackage{amsmath}	

\usepackage{amssymb}	
\usepackage{mathtools}  
\numberwithin{equation}{section}    
\usepackage{booktabs}   
\usepackage{longtable}
\usepackage{supertabular}
\usepackage{pdflscape} 

\DeclarePairedDelimiterX{\norm}[1]{\lVert}{\rVert}{#1}

\title[Size and stellar mass evolution of satellite ETGs]{Newcomers and suburbanites can drive the evolution of the size-stellar mass relation of early type galaxies in galaxy clusters}

\author[Matteuzzi et al.]{
Massimiliano Matteuzzi,$^{1,2}$\thanks{E-mail: massimilia.matteuzz2@unibo.it}
Federico Marinacci,$^{1}$
Carlo Nipoti,$^{1}$
Stefano Andreon$^{3}$\\
$^{1}$Department of Physics \& Astronomy "Augusto Righi", University of Bologna, via Gobetti 93/2, 40129 Bologna, Italy\\
$^{2}$INAF-Astrophysics and Space Science Observatory of Bologna, via Gobetti 93/3, 40129 Bologna, Italy\\
$^{3}$INAF–Osservatorio Astronomico di Brera, via Brera 28, 20121, Milano, Italy\\
}

\pubyear{2022}

\begin{document}
\label{firstpage}
\pagerange{\pageref{firstpage}--\pageref{lastpage}}
\maketitle

\begin{abstract}
At fixed stellar mass $M_*$, the effective radius $R_{\rm e}$ of massive satellite early-type galaxies (ETGs) in galaxy clusters is, on average, larger at lower redshift. We study theoretically this size evolution using the state-of-the-art cosmological simulation IllustrisTNG100: we sampled $75$ simulated satellite ETGs at redshift $z=0$ with $M_* \ge 10^{10.4} M_{\sun}$ belonging to the two most massive ($\approx 10^{14.6} M_{\sun} $) haloes of the simulation. We traced back in time the two clusters’ main progenitors and we selected their satellite ETGs at $z>0$ with the same criterion adopted at $z=0$. The $R_{\rm e}-M_*$ relation of the simulated cluster satellite ETGs, which is robustly measured out to $z=0.85$, evolves similarly to the observed relation over the redshift range $0\lesssim z \lesssim 0.85$. In the simulation the main drivers of this evolution  are the acquisition of new galaxies ("newcomers") by the clusters and the transformation of member galaxies located at large clustercentric distance ("suburbanites") at $z=0.85$, which end up being massive satellite ETGs at $z=0$. Though several physical processes contribute to change the population of satellite ETGs in the considered redshift interval, the shape of the stellar mass function of the simulated cluster ETGs is not significantly different at $z=0.85$ and at $z=0$, consistent with observations.
\end{abstract}




\section{Introduction}
\label{sec:intro}
The evolution of galaxies is expected to be significantly influenced by the environment in which they are located. This paper focuses on the evolution of galaxies belonging to clusters of galaxies: in this extreme, high-density environment galaxies are subjected to several physical processes that can affect their evolution. Dynamical friction \citep[e.g.][]{1943ApJ....97..255C,2021ApJ...916...55T} exerted on the most massive galaxies favours their infall towards the centre and, consequently, the merger with the central brightest cluster galaxy \citep[BCG; e.g.][]{1978ApJ...224..320H,2017MNRAS.467..661N}. In addition, harassment \citep[e.g.][]{1976ApJ...209..382L,2015A&A...576A.103B}, ram pressure stripping \citep[e.g.][]{1972ApJ...176....1G,2020MNRAS.496.2673J} and tidal stripping \citep[e.g.][]{2006AJ....131.2417C,2020A&A...638A.133L} may transform cluster disc galaxies into early-type galaxies (ETGs) with very low specific star formation rate, lower stellar mass and smaller size than the ones they had before being acquired by the cluster. This is consistent with the finding that cluster late-type galaxies (LTGs) have less gas than their field counterparts and with the Butcher-Oemler effect on dwarf galaxies \citep[e.g.][]{2003ApJ...598...20D,2012A&A...537A..88R,2013MNRAS.434.3469D}: clusters at intermediate redshifts ($z \approx 0.5$) have many blue dwarf galaxies and a few red dwarf ellipticals in the central zones, whereas the central parts of nearby clusters ($z<0.1$) have  many red dwarf ellipticals and a few blue dwarf galaxies. Moreover, Hubble space telescope (\emph{HST}) observations \citep{1994ApJ...430..107D,1994ApJ...430..121C} found that these blue dwarf galaxies are disturbed spirals that experienced many star formation events spaced 1-2 Gyr apart. Therefore, satellite galaxies in clusters, and especially galaxies that are very extended, are severely affected by the accumulation of galaxy encounters.

The combination of the aforementioned processes is believed to determine the difference between the observed properties of cluster and field galaxies. Not only the mixture of morphological types, but also the average properties of the galaxies of a given morphological type are found to depend on the environment \citep[e.g.][]{1976ApJ...206..883V,1984ARA&A..22..445H,1985ApJ...292..404G}. In particular, here we consider cluster ETGs, addressing the question of how the evolution of their sizes and stellar masses are influenced by the cluster environment. Present-day cluster ETGs are in general found to follow a correlation between size and stellar mass similar to that of present-day field ETGs  \citep[e.g.][]{1998AJ....116.1606P,2013ApJ...778L...2C,2014MNRAS.444..682C}, though evidence of non-negligible environmental dependence of the $z\approx 0$ size-stellar mass relation has been reported by some authors \citep[e.g.][]{2010ApJ...712..226V,2017ApJ...834...73Y}.
ETGs in higher redshift clusters tend to be more compact than their $z\approx 0$ counterparts: 
they also follow a size-stellar mass relation, but such that, for given stellar mass, the average size decreases for increasing $z$  \citep[e.g.][]{2012ApJ...745..130R,2016A&A...593A...2A,2021MNRAS.507.5272N}. 
However, it is debated whether, at these higher redshifts (say $z\gtrsim 1$), cluster ETGs are, on average, more or less extended than field ETGs of the same stellar mass \citep[][]{2012ApJ...745..130R,2013ApJ...770...58B,2014ApJ...788...51N,2018A&A...617A..53A,2018ApJ...856....8C,2019MNRAS.484..595M,2021MNRAS.507.5272N}. The variety of these results reflects the difficulty in comparing different data with a correct estimate of the systematic errors. For instance, half-light radii are sometimes computed assuming that galaxies have a single stellar component and/or that all isophotes have the same ellipticity and position angle, which in general is not justified \citep[e.g.][]{2011ApJ...730...38V,2014ApJ...788...11L}. Moreover, the selection of ETGs is often heterogeneous in different works, which use different criteria based on morphology, star formation rate and/or colours. For instance, UVJ-selected galaxies are a mix of at least three populations plausibly evolving in different ways \citep[see e.g.][]{2009ApJ...691.1879W,2013ApJ...773..112C,2013A&A...558A..61M,2020A&A...640A..34A}.

In this paper we study from a theoretical point of view the evolution of massive satellite ETGs (i.e.\ non BCGs) in clusters of galaxies, taking as reference the observational dataset of \citet{2016A&A...593A...2A}, in which ETGs are selected and analysed in a homogeneous way in clusters at different redshifts. \citet{2016A&A...593A...2A} measured the effective (i.e. half-light) radius $R_{\rm e}$ and the stellar mass $M_*$ of satellite ETGs in 14 massive clusters (virial mass $\approx 10^{14.2-15.3} M_\odot$) in the redshift range $0.02 \le z \le 1.80$, finding that ETGs of given $M_*$ have, on average, larger $R_{\rm e}$ at lower $z$. Similar results are found when a radius that encloses $80\%$ of the total light is used instead of $R_{\rm e }$ \citep{2020A&A...640A..34A}. ETGs at $1.4 \lesssim z \lesssim 3 $ have, for a given $M_*$, a size much smaller than their today's counterparts \citep[their effective radius can be less than 1 kpc, see e.g.][and references therein]{2005ApJ...626..680D,2014ApJ...788...28V}.
The findings of \citet{2016A&A...593A...2A} were compared by \citet{2018A&A...617A..53A} with field red sequence ETGs selected in the same way: the size growth of cluster galaxies is much weaker than that of the field galaxies. For example, at $z=1.80$ cluster ETGs with $M_* = 10^{11}M_\odot$ have, on average, $R_{\rm e} \approx 1.2$ kpc, which is $\approx 1.6$ times the size of their field counterparts, but at $z=0$ the corresponding cluster ETGs have $R_{\rm e}\approx 2.1$ kpc, i.e. $\approx 94\%$ of the size of their field counterparts (\citeauthor{2016A&A...593A...2A} \citeyear{2016A&A...593A...2A}; see also \citeauthor{2018A&A...617A..53A} \citeyear{2018A&A...617A..53A}). The same conclusion is also obtained by combining \citet{2016A&A...593A...2A} data with the field quiescent galaxies observed by \citet{2012ApJ...746..162N}.

The growth of field ETGs is usually explained as mainly due to dry mergers \citep[e.g.][]{2009ApJ...699L.178N,2012MNRAS.422.1714N}, but this mechanism is not expected to be efficient in a cluster environment because its high velocity dispersion disfavour galaxy mergers \citep[see e.g.][\S 8.9]{cimatti2019introduction}, of course with the exception of the dynamical friction-driven mergers with the central BCG. Thus, the observed size growth of the population of cluster satellite ETGs, though weaker than that of field ETGs, poses a challenge to theoretical models of galaxy evolution. In this work we address this question using results from the IllustrisTNG100 cosmological simulation \citep{2018MNRAS.480.5113M,2018MNRAS.477.1206N,2018MNRAS.475..624N,2018MNRAS.475..648P,2018MNRAS.475..676S,2019ComAC...6....2N}. In particular, we will compare the distributions in the $M_*-R_{\rm e}$ plane of simulated and observed ETGs at different redshifts, and we will follow in time individual simulated galaxies to shed light on the overall evolution of the population of cluster satellite ETGs.

The paper is organized as follows. We describe the sample of observed galaxies in Section~\ref{sec:obs} and the sample of simulated galaxies in Section~\ref{sec:anal}. In Section~\ref{sec:results} we present the results obtained comparing the observed and simulated galaxy samples. Section~\ref{sec:concl} concludes.

\section{Observational data}
\label{sec:obs}
The reference observational data that we consider in this work are the measurements of $R_{\rm e}$ and $M_*$ of satellite ETGs in clusters of galaxies obtained by \citet{2016A&A...593A...2A}. Here, we briefly describes how these quantities are derived from observations.

In \citet{2016A&A...593A...2A} $R_{\rm e}$ and $M_*$ of galaxies are determined with an isophotal analysis carried out in the observed band closest to the rest-frame $r$ band, i.e. F625W, F814W or F850LP filters for clusters at $z<1$. Each isophote is fitted with an ellipse plus deviations from a perfect ellipse. A growth curve (i.e. the projected cumulative stellar light of a galaxy) is computed with an analytical integration of the flux between isophotes up to the last one detected and an extrapolation to infinity is then performed by fitting a library of growth curves of nearby ETGs \citep[][]{1977ApJS...33..211D}. The background light is accounted for via low-order polynomial fitting: the final result is the corrected total galaxy flux. Finally, $M_*$ is derived from the flux at $\approx 600$ nm assuming a single stellar population (SSP) computed with the stellar population synthesis model of \citeauthor{2003MNRAS.344.1000B} (\citeyear{2003MNRAS.344.1000B}; BC03) formed at $z=3$ with solar metallicity, \citet{1955ApJ...121..161S} initial mass function (IMF) and cosmological parameters $h=0.7$, $\Omega_{\rm m} =0.3$, $\Omega_{\rm \Lambda} =0.7$. The values of these cosmological parameters adopted in the IllustrisTNG100 simulation are slightly different (Section \ref{sec:simul}): this means that $R_{\rm e}$ and $M_*$ calculated in the simulation are higher than in observations by $\approx 0.014$ dex. However, the intrinsic scatter of the $R_{\rm e}-M_*$ relations (Section \ref{sec:results}) and the systematic errors in the $M_*$ measure are much higher than the difference due to the cosmological parameters: from now on we will neglect this contribution.
\citet{2016A&A...593A...2A} selected only ETGs that are satellites with $M_* \ge 10^{10.65} M_\odot$ (Salpeter IMF) and characterise the size of a galaxy by measuring its half-light circularized radius $R_{\rm e}=\sqrt{A_{\rm e}/\pi}$, where $A_{\rm e}$ is the area of the surface within the isophote enclosing half the total light.

\section{Analysis of the simulation}
\label{sec:anal}
In this section we describe the approach that we adopted to analyse the simulation TNG100 of the IllustrisTNG simulation suite. In particular, we focus on the description of the TNG100 simulation (Section \ref{sec:simul}), the morphological classification of the simulated galaxies (Section \ref{sec:class}), the selection of our sample of simulated satellite ETGs (Section \ref{sec:selec}), and the computation of their stellar mass and effective radius (Section \ref{growthcurve}).

\subsection{The TNG100 simulation}
\label{sec:simul}
We use the highest resolution realisation of the cosmological simulation TNG100, one of the runs of the IllustrisTNG project whose data are publicly available. The simulation adopts the \citet{2016A&A...594A..13P} cosmological parameters ($h=0.6774$, $\Omega_{\rm m} =0.3089$, $\Omega_{\rm \Lambda} =0.6911$) and it has been run with the moving-mesh code {\sc arepo} \citep{2010MNRAS.401..791S} that solves ideal magnetohydrodynamics coupled with self-gravity. TNG100 follows the evolution of $2 \times 1820^3$ resolution elements over time (from $z = 127$ to $z = 0$) within a periodic cube of $75h^{-1} \text{Mpc} \simeq 110.7$ Mpc on a side at $z=0$, large enough to include massive clusters. The average mass per baryonic particle is $ 9.4 \cdot 10^{5}h^{-1}M_\odot \simeq 1.4 \cdot 10^{6} M_\odot$ and the $z= 0$ gravitational (Plummer equivalent) softening length for stellar particles is $0.5 h^{-1} \text{kpc} \simeq 0.74$ kpc. The IllustrisTNG suite studies the evolution of dark matter (DM), gas, stars and supermassive black holes, and it implements prescriptions for physical processes that play a key role for galaxy formation and evolution such as microphysical gas radiative mechanisms, star formation in the dense interstellar medium, stellar population evolution and chemical enrichment, stellar feedback and black hole feedback operating in a thermal "quasar" mode at high accretion states and a kinetic "wind" mode at low accretion states \citep[see][for a detailed description of the IllustrisTNG model]{2017MNRAS.465.3291W,2018MNRAS.473.4077P}.

In TNG100, haloes are identified with a friends-of-friends (FoF) algorithm \citep{1985ApJ...292..371D} that has a linking length $b=0.2$ in units of the mean interparticle distance. This algorithm is run on DM particles that are linked together to the same structure if their distance is less than the linking length. Other particle types (gas, stars and black holes) are assigned to the same FoF group as their closest DM particle (the search for this particle is limited to 4 times the linking length). In the resulting group catalogue (and in the snapshots as well), structures are listed in decreasing order of total mass at a selected time and with an index starting from 0: for example, the most massive structure, called "FoF 0", at $z=0$ might not be the most massive at higher redshifts \citep[see][for details]{2019ComAC...6....2N}. In this work we analyse the $z=0$ two most massive galaxy clusters (i.e. "FoF 0" and "FoF 1" structures). 
Each halo includes subhaloes that are gravitationally bound structures with more than 20 particles. In the simulations and in this work we identify each subhalo with a single galaxy through the {\sc subfind} algorithm \citep{2001MNRAS.328..726S}, which is able to detect hierarchies of the substructure. Thanks to {\sc subfind}, we can use just an index (ID) to identify each subhalo in the catalogue. We chose only subhaloes formed by the processes of structure formation and collapse, i.e. subhaloes of cosmological origin identified by "SubhaloFlag = True" (for more details see \citealt{2019ComAC...6....2N}).

Actually, the FoF algorithm cited so far could lead to biased final results, because the physical linking length increases with lowering redshift (the mean interparticle distance is in comoving units). This means that at $z=0$ a halo identified with the FoF algorithm comprises the whole cluster, whereas at high redshift it only includes the central regions of a cluster. In Section \ref{sec:membership} we will investigate possible biases by adopting a more inclusive membership criterion with increasing redshift.

\subsection{Classification of galaxies} 
\label{sec:class}
Several classification methods have been used to classify galaxies both in observations and in simulations \citep[e.g.][]{2004AJ....128..163L,2019MNRAS.483.4140R,2019MNRAS.487.5416T}. \citet{2016A&A...593A...2A} classify the morphology of real galaxies by their radial profiles of the isophote parameters: bars, discs, bulges, spiral arms and \ion{H}{ii} regions have distinctive signatures \citep[see also][]{1977ApJS...33..211D}; for example, spiral arms and other irregularities are measured by deviations of the isophote from ellipses.
Ideally, one would like to use the same classification method also for the simulated galaxies, but this is in general not possible in galaxies generated in cosmological simulations because of their limited spatial resolution. The resolution of the TNG100 simulation is sufficiently high to compute reliably the size, stellar mass and global kinematics of the massive galaxies of our sample, but not to determine local properties such as the radial profiles of the isophote parameters. In particular, we cannot use the isophotal method, because the softening length for stellar particles (Section \ref{sec:simul}) is greater than the minimum spatial resolution needed to interpret the data correctly \citep[][]{1980ApJS...42..565D,1997A&A...319..747A}.  Therefore, similarly to \citet{2020MNRAS.496.2673J} and \citet{2020A&A...638A.133L}, we classify galaxies on the basis of global stellar kinematic properties, i.e.\ with the rotation parameter $f$ defined as the stellar mass fraction with circularity parameter larger than 0.7 \citep[see e.g.][]{2014MNRAS.437.1750M}. The circularity parameter is defined as the ratio $J_z/J(E)$, where $J_z$ is the specific angular momentum component of a given stellar particle along the galaxy spin axis and $J(E)$ is the maximum specific angular momentum magnitude of the stellar particles at positions between 50 before and 50 after the particle under examination in a list where the stellar particles are sorted by their binding energy \citep{2015ApJ...804L..40G}. We define ETGs as the subhaloes with $f<0.3$ calculated including all stars in the subhalo: LTGs are those with $f \ge 0.3$. We expect that our selection using $f$ extracts a population morphologically similar to that of \citeauthor{2016A&A...593A...2A} (\citeyear{2016A&A...593A...2A}; see \citeauthor{2019MNRAS.483.4140R} \citeyear{2019MNRAS.483.4140R}; \citeauthor{2020MNRAS.496.2673J} \citeyear{2020MNRAS.496.2673J}; \citeauthor{2020A&A...638A.133L} \citeyear{2020A&A...638A.133L}).

\subsection{Sample selection}
\label{sec:selec}
Thanks to the {\sc SubLink} algorithm \citep{2015MNRAS.449...49R}, we can determine which galaxies belong to each cluster in any snapshot (i.e. at any sampled redshift).
In each of the two simulated clusters at $z=0$, we selected galaxies with $M_* \ge 10^{10.4} M_\odot$ that are satellite (i.e. we exclude the central galaxy of each cluster) and that are classified as ETG according to the criterion described in Section \ref{sec:class}. At higher redshifts, we tracked the main progenitor of each of the two clusters and we selected satellite ETGs using the same criterion as at $z=0$. In Tables~\ref{tab:Mhalf} - \ref{tab:Mhalf54444} we list for our sample of ETGs {\sc subfind} IDs, $M_*$ and $R_{\rm e}$, calculated with the growth curve method described in Section~\ref{growthcurve}, at $z=0$ and $z=0.85$. We take $z=0.85$ as the highest-redshift snapshot, because at higher redshifts we do not have enough statistics to infer robustly the properties of the size-stellar mass relation (see Appendix \ref{app:evzhigh}). The $z=0$ sample consists of 75 ETGs, whereas the $z=0.85$ sample is composed by 29 ETGs.

\subsection{Growth curve}
\label{growthcurve}
We define the growth curve $M(R)$ of a simulated galaxy as its projected cumulative distribution of stellar mass calculated within a given projected radius $R$: projected properties are defined using a single line of sight, along the $z$-axis of the TNG100 simulation box. We use the growth curve $M(R)$ to calculate the corresponding projected stellar half-mass radius for a given total stellar mass definition: assuming that the total stellar mass is the one within twice the 3D stellar half-mass radius $M_{*,2r_{\rm h}}$, we compute the projected stellar half-mass radius $R_{{\rm e},2r_{\rm h}}$ such that $M(R_{{\rm e},2r_{\rm h}}) = M_{*,2r_{\rm h}}/2$. We refer the reader to Appendix \ref{app:massdef} for a justification of this choice and for the definition of $r_{\rm h}$. For the rest of this paper, for simulated galaxies, we will identify the effective radius $R_{\rm e}$ with $R_{{\rm e},2r_{\rm h}}$ and the stellar mass $M_*$ with $M_{*,2r_{\rm h}}$.

\section{Results}
\label{sec:results}
In this section we present the results that we obtained studying the evolution of the $R_{\rm e}-M_*$ relation from $z=0.85$ to $z=0$ (Section \ref{sec:chap3}), what causes the evolution of the relation with redshift (Section \ref{sec:evoindivual}), the dependence of the results on different membership definitions (Section \ref{sec:membership}), and the evolution of the stellar mass function of cluster ETGs (Section \ref{sec:StellMassFunc}).

\subsection{Evolution of the size-stellar mass relation}
\label{sec:chap3}
Using the quantities introduced in Section~\ref{sec:anal}, we can populate the stellar mass-size plane with our sample of simulated ETGs, compute the $R_{\rm e}-M_*$ relations at $z=0$ and $z=0.85$, and compare them with the corresponding observed relations. The stellar masses of \citet{2016A&A...593A...2A}, derived using \citet{1955ApJ...121..161S} IMF, are converted into stellar masses for \citet{2003PASP..115..763C} IMF, used in TNG100, simply subtracting 0.25 dex \citep[e.g.][]{2010ApJ...724..511A,2017MNRAS.465.2397S}. With this conversion, the stellar mass cut of Andreon et al.\ (\citeyear{2016A&A...593A...2A}, \citeyear{2020A&A...640A..34A}) is the same ($M_* \geq 10^{10.4} M_\odot$; Chabrier IMF) as that adopted for our sample of simulated ETGs.

\subsubsection{Size-stellar mass relation at \texorpdfstring{$z$ = 0}{z = 0}}
\label{sec:Sizez0}
\begin{figure}
\includegraphics[width=\columnwidth]{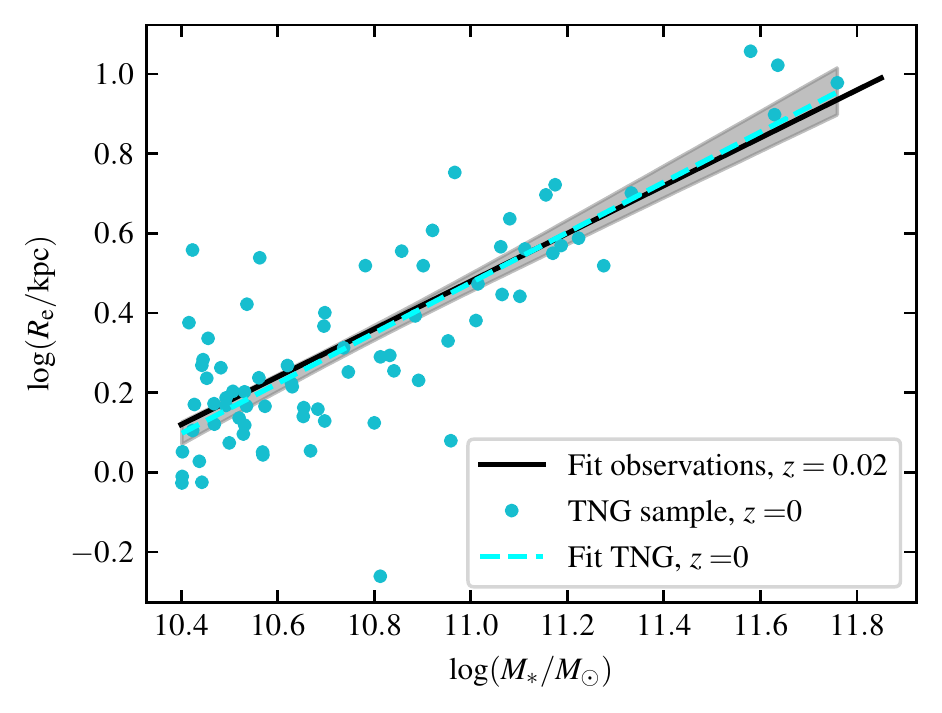}
\caption{Effective radius $R_{\rm e}$ as a function of stellar mass $M_*$ at $z=0$ for our sample of TNG ETGs (filled cyan circles). Their median size-stellar mass relation (cyan dashed line) along with its $68\%$ uncertainty (posterior highest density interval as a grey band) are also presented. The black solid line represents \citet{2016A&A...593A...2A} best fit to the Coma cluster ETGs. The dashed line is the same, within the errors, as the solid one.}
\label{fig:s-m1}
\end{figure}

In Fig.~\ref{fig:s-m1} we show the effective radius $R_{\rm e}$ as a function of the total stellar mass $M_*$ for the simulated ETGs of our sample at $z=0$. We fit the linear relation $\log (R_{\rm e}/\mbox{kpc})=\alpha \left[ \log (M_*/M_\odot) - 10.75 \right]+\gamma$ with an intrinsic scatter $\sigma$. We assume that the uncertainties on $M_*$ and $R_{\rm e}$ are negligible for the simulated galaxies, which are modelled with a sufficiently high number of stellar particles (in the range\footnote{This is calculated considering a $M_*$ range of $10^{10.4} - 10^{11.76} M_\odot$ and an average mass per baryonic particle of $\approx 1.4 \cdot 10^{6} M_\odot$} $ \approx 1.8 \times 10^{4} - 4.1 \times 10^{5} $). We performed a Bayesian analysis for the fitting procedure, adopting uniform priors for $\gamma$, $\sigma$ and $\arctan \alpha$ \citep[the latter following][]{2010MNRAS.404.1922A}. The median size-stellar mass relation is represented in Fig.~\ref{fig:s-m1} along with its $68\%$ uncertainty (posterior highest density interval) as a cyan dashed line and a grey band respectively, and nicely matches the relation found by \citet{2016A&A...593A...2A} for the $z=0.02$ Coma cluster ETGs (black solid line derived in the same way). 
The simulated $R_{\rm e}-M_*$ relation has the same intrinsic scatter $\sigma\simeq 0.16$ as the observed relation. All fit parameters and errors are listed in Table~\ref{tab:Pearson54}.

\subsubsection{Size-stellar mass relation at \texorpdfstring{$z$ = 0.85}{z = 0.85}}
\label{sec:sizehighz}
\begin{figure}
\includegraphics[width=\columnwidth]{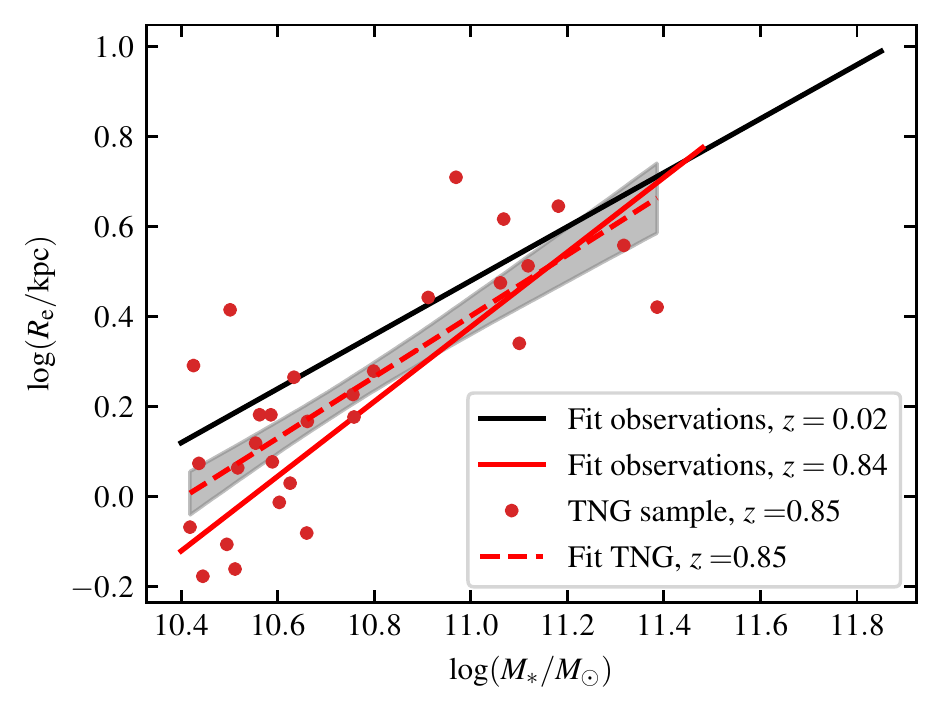}
\caption{Effective radius $R_{\rm e}$ as a function of stellar mass $M_*$ at $z=0.85$ for our sample of TNG ETGs (filled red circles). Their median size-stellar mass relation (red dashed line) along with its $68\%$ uncertainty (posterior highest density interval as a grey band) are also shown. The black and red solid lines represent \citet{2016A&A...593A...2A} best fit to the Coma cluster ($z=0.02$) ETGs and to the RXJ0152.7-1357 cluster ($z=0.84$) ETGs, respectively. The red dashed line is similar to the red solid one: they have a lower normalisation than the black solid line.}
\label{fig:s-mall}
\end{figure}

\citet{2016A&A...593A...2A} studied 14 clusters, measuring the size-stellar mass relation of satellite ETGs in different redshift bins: in particular, here we consider observational data at $z=0.84$ (RXJ0152.7-1357 cluster), comparing them with data from the $z=0.85$ snapshot of TNG100.
We build our sample of simulated cluster ETGs at $z=0.85$ applying the same selection criteria as at $z=0$ (see Section \ref{sec:selec}).
Figure \ref{fig:s-mall} shows the median size-stellar mass relation at $z=0.85$ (red dashed line), whereas the black and red solid lines represent \citet{2016A&A...593A...2A} best fits to the Coma cluster ETGs ($z=0.02$) and to the RXJ0152.7-1357 cluster ETGs ($z=0.84$), respectively.
All fit parameters and errors are listed in Table~\ref{tab:Pearson54}. The simulated galaxies follow a trend that is similar to that of the observed ones: the simulated $R_{\rm e}-M_*$ relation at $z=0.85$ has the same slope and intercept, within the errors, as the observed relation at $z=0.84$. Moreover, the simulated $R_{\rm e}-M_*$ relation at $z=0.85$ has similar slope, but lower normalisation than at $z=0$ (higher-redshift galaxies have smaller $R_{\rm e}$ for given $M_*$): the difference between the two $R_{\rm e}-M_*$ relations is $\approx 3 \sigma$. This is slightly smaller than the difference between the observed $R_{\rm e}-M_*$ relation at $z=0.84$ and $z=0.02$, which is $\approx 4 \sigma$. Remarkably, the intrinsic scatter of the simulated $R_{\rm e}-M_*$ relation at $z=0.85$ is the same, within the errors, as that of the $z=0$ relation, consistent with what found for the observational sample (see Table~\ref{tab:Pearson54}). Appendix \ref{app:evzhigh} shows that an agreement between observations and simulations is found also at $z=1.3$, where, however, the sample of simulated ETGs is too small to draw robust conclusions.
\begin{table*}
\centering
\caption{The redshift $z$, mean slope $\alpha$, intercept $\gamma$ and intrinsic scatter $\sigma$ of the linear relation $\log (R_{\rm e}/\mbox{kpc})=\alpha \left[ \log (M_*/M_\odot) - 10.75 \right]+\gamma$, along with their errors, of the best-fitting relations displayed in Figures \ref{fig:s-m1} - \ref{fig:membership}. Here $N$ indicates the number of galaxies in each subsample.}
\label{tab:Pearson54}
\begin{tabular}{@{}llllll@{}}
\toprule
Figure  & Fig. 1, 2 fit observations & Fig. 1, 3, 5 fit TNG & Fig. 5, remain (fit) & Fig. 3, 4 acquired (fit) & Fig. 4, ETGs acquired (fit) \\\midrule
$z$            & 0.02                  & 0              & 0        & 0 & 0 \\
$\alpha$       & $0.60 \pm 0.06$       & $0.63 \pm 0.06$   & $0.50 \pm 0.17$    & $0.61 \pm 0.07$ & $0.67 \pm 0.11$ \\
$\gamma$       & $0.33 \pm 0.02$       & $0.318 \pm 0.019$   & $0.25 \pm 0.04$  & $0.34 \pm 0.03$  & $0.30 \pm 0.05$ \\
$\sigma$       & $0.16 \pm 0.01$       & $0.161 \pm 0.014$  & $0.08 \pm 0.03$   & $0.180 \pm 0.018$  & $0.21 \pm 0.03$ \\ 
$N$            & 86                    & 75        & 8              & 53 & 27 \\ \bottomrule
\toprule
Figure  & Fig. 2, fit observations &   Fig. 2, 3, 5, 6 fit TNG & Fig. 5, remain (fit) & Fig. 4, LTGs acquired (fit)  & Fig. 6, no merger (fit) \\ \midrule
$z$             & 0.84 & 0.85 &  0.85 &  0  &  0.85 \\
$\alpha$       & $0.83 \pm 0.12$ & $0.68 \pm 0.11$ & $0.7 \pm 0.3$ & $0.61 \pm 0.13$  & $0.74 \pm 0.19$ \\
$\gamma$       & $0.17 \pm 0.04$ &  $0.23 \pm 0.03$ & $0.21 \pm 0.07$ & $0.37 \pm 0.03$  & $0.28 \pm 0.04$ \\
$\sigma$       & $0.18 \pm 0.02$ &  $0.17 \pm 0.02$ & $0.17 \pm 0.07$ & $0.15 \pm 0.02$  & $0.17 \pm 0.03$ \\
$N$            &  21    & 29 & 8 & 26  & 9 \\ \bottomrule
\toprule
Figure  & Fig. 7, acquired (fit)  & Fig. 7, included (fit) & Fig. 7, remain (fit)\\ \cline{1-4}
$z$            & 0 & 0 & 0\\
$\alpha$       & $0.64 \pm 0.13$ & $0.59 \pm 0.10$ & $0.58 \pm 0.10$\\
$\gamma$       & $0.33 \pm 0.04$ & $0.35 \pm 0.04$ & $0.27 \pm 0.03$\\
$\sigma$       & $0.19 \pm 0.03$ & $0.19 \pm 0.03$ & $0.104 \pm 0.018$\\
$N$            & 27 & 26 & 22\\ \cline{1-4}\cline{1-4}
\end{tabular}
\end{table*}

\subsection{Evolution of individual galaxies}
\label{sec:evoindivual}
Once ascertained that TNG100 reproduces the observed evolution of the $R_{\rm e}-M_*$ relation of cluster ETGs, we would like to understand what causes the change of the size-stellar mass relation between $z=0.85$ and $z=0$. This can be done by studying the individual evolution of our simulated clusters' galaxies. Via the {\sc SubLink} algorithm, we followed the main progenitors of all the ETGs of our sample from $z=0$ back to $z=0.85$, and we followed the main descendants of all the ETGs of our $z=0.85$ sample from $z=0.85$ to $z=0$ (see Section \ref{sec:mergeBCG}).
We observe different evolutionary paths, which we describe separately in the following subsections.

Only a minority (12 out of 75) of the galaxies that belong to our $z=0$ sample of simulated satellite ETGs also belong to our $z=0.85$ sample. Of these 12, 8 are ETGs that never change their morphology between $z=0.85$ and $z=0$, while 4 do (i.e.\ these 4 galaxies become LTGs at $0<z<0.85$ and then end up being ETGs at $z=0$).
Other 10 galaxies belonged to the clusters at $z=0.85$, but are classified as LTGs at $z=0.85$.
Most of the $z=0$ galaxies (53 out of 75) have been acquired by the clusters in the redshift interval $0<z<0.85$. We thus start our analysis by considering this population of acquired galaxies.

\subsubsection{Galaxies that are acquired by the clusters}
\label{sec:resacq}
\begin{figure}
\includegraphics[width=\columnwidth]{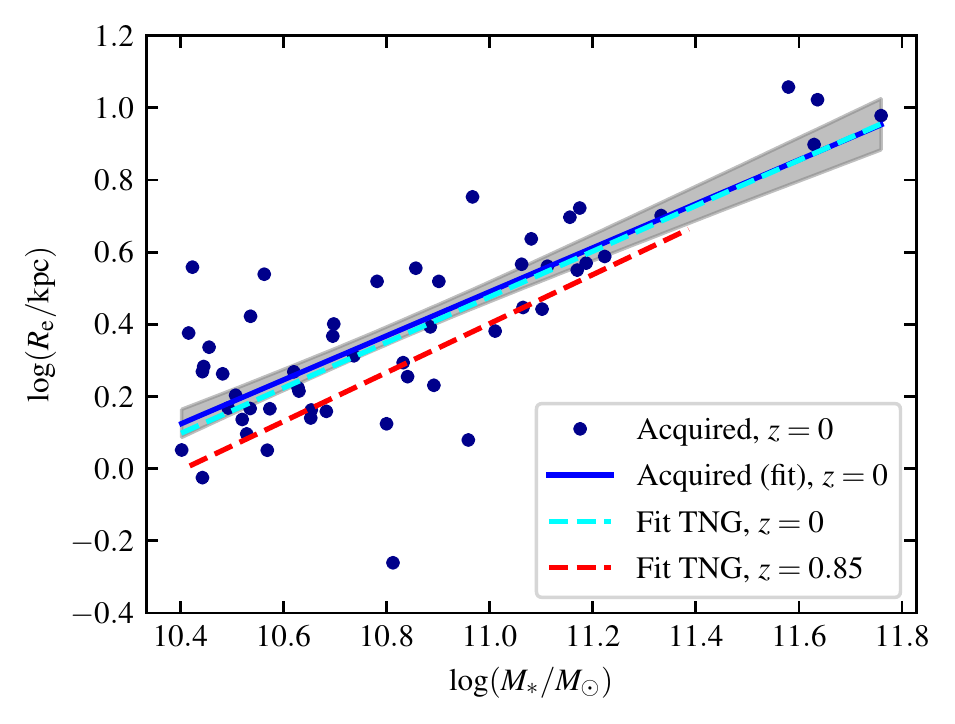}
\caption{Effective radius $R_{\rm e}$ as a function of stellar mass $M_*$ at $z=0$ for the 53 satellite ETGs of the simulated clusters acquired at $z<0.85$ (filled blue circles). The median size-stellar mass relation (solid blue line) is shown along with its $68\%$ uncertainty (posterior highest density interval as a grey band). For comparison, the two dashed lines are the median $R_{\rm e}-M_*$ relations for all the satellite ETGs of the clusters at $z=0.85$ (red) and $z=0$ (cyan). The solid line is the same, within the errors, as the cyan dashed one: they have a higher normalisation than the red dashed line.}
\label{fig:s-mev6}
\end{figure}

\begin{figure}
\includegraphics[width=\columnwidth]{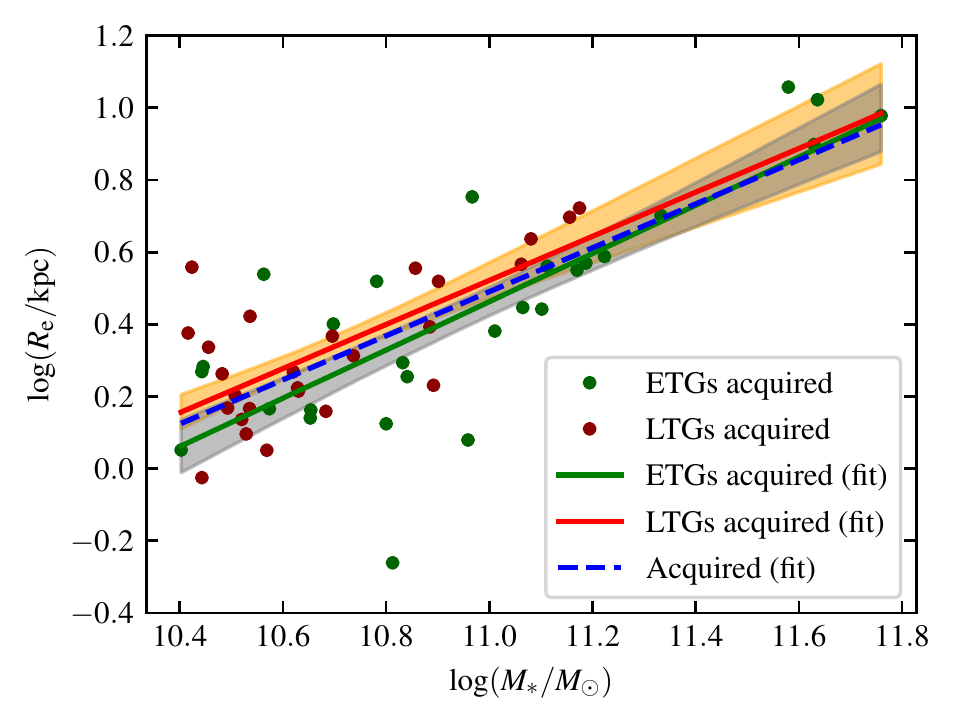}
\caption{Effective radius $R_{\rm e}$ as a function of stellar mass $M_*$ at $z=0$ for the 53 satellite ETGs of the simulated clusters acquired at $z<0.85$. The green circles are the 27 galaxies that were ETGs also at $z=0.85$, whereas the red ones are the 26 galaxies that were LTGs at $z=0.85$. The two median size-stellar mass relations (solid green and red lines) are shown along with their $68\%$ uncertainty (posterior highest density interval as a grey and orange band respectively). For comparison, the dashed blue line is the median $R_{\rm e}-M_*$ relation for all the 53 acquired satellite ETGs at $z=0$. The green and red solid lines are the same, within the errors, as the blue dashed one.}
\label{fig:s-mev6ETGsLTGs}
\end{figure}

Figure~\ref{fig:s-mev6} shows the effect of the galaxies acquired by the clusters at $z<0.85$ on the size-stellar mass relation evolution: this plot suggests that this evolution can be explained mainly thanks to these newly acquired galaxies. The median $R_{\rm e}-M_*$ relation of these 53 $z=0$ ETGs (solid blue line) is the same, within the errors, as the median relation of the full sample at $z=0$ (dashed blue line). These acquired galaxies tend to be more extended than the ETGs that already belonged to the two clusters. A possible explanation for this might be the different evolutionary history that acquired galaxies have compared to cluster galaxies.
Some of the acquired galaxies underwent morphological transformations since $z=0.85$. We thus consider separately those that were ETGs also at $z=0.85$ (27 galaxies; green in Fig. \ref{fig:s-mev6ETGsLTGs}) and those that were LTGs at $z=0.85$ (26 galaxies; red in Fig. \ref{fig:s-mev6ETGsLTGs}).
Fig. \ref{fig:s-mev6ETGsLTGs} shows that, at $z=0$, the most massive ($M_* > 10^{11.2} M_\odot $) acquired galaxies were ETGs at $z=0.85$, while $\approx 55\%$ of the acquired galaxies with $M_* < 10^{11.2} M_\odot $ were LTGs at $z=0.85$.
Table~\ref{tab:Pearson54} shows the best fit parameters and errors of the two aforementioned subsamples: the $z=0$ best fits of the acquired ETGs and LTGs are indistinguishable, which suggests that both families of acquired galaxies contribute to the evolution of the $R_{\rm e}-M_*$ relation, independent of their morphological transformations since $z=0.85$.

\subsubsection{ETGs that remain satellites of the cluster}
\begin{figure}
\includegraphics[width=\columnwidth]{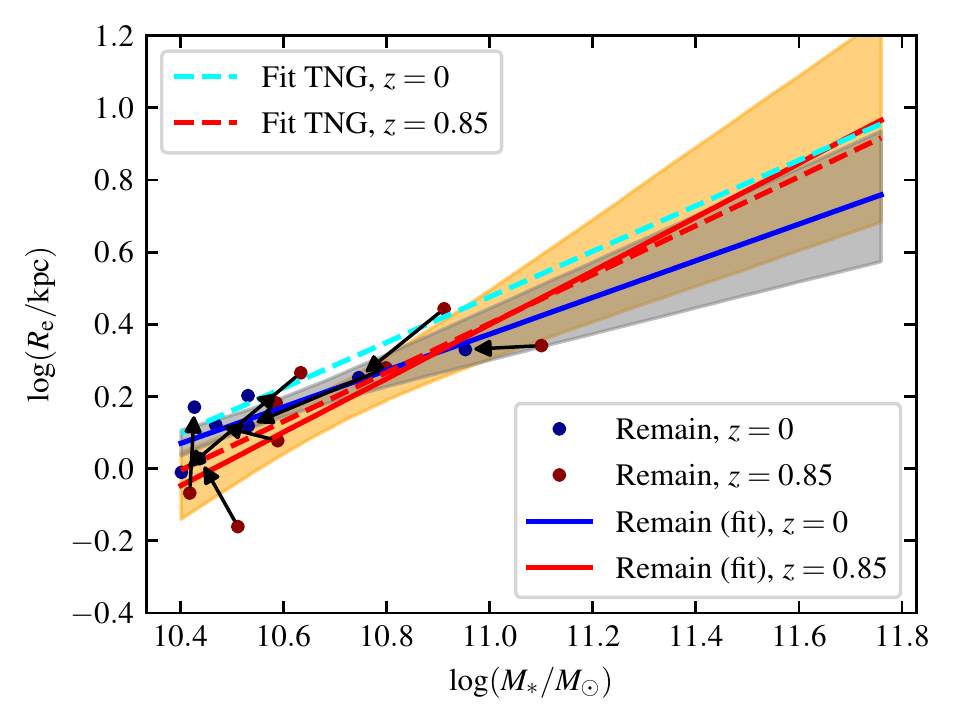}
\caption{Effective radius $R_{\rm e}$ as a function of stellar mass $M_*$ for the 8 satellite ETGs of the simulated clusters that remain satellite ETGs of the same cluster from $z=0.85$ (red circles) to $z=0$ (blue circles): the arrows connect the points representing the same galaxy. The two median size-stellar mass relations at $z=0$ and $z=0.85$ of these 8 ETGs are solid red and blue lines, whereas the colored bands are their $68\%$ uncertainties (posterior highest density interval) at $z=0$ (grey) and $z=0.85$ (orange). For comparison, the median size-stellar mass relations for the entire sample of satellite ETGs of the clusters at $z=0.85$ and $z=0$ are shown as dashed lines in red and cyan respectively. The red dashed line is the same, within the errors, as the red and blue solid lines.}
\label{fig:s-mev1}
\end{figure}

Here we focus on the evolution of the galaxies that were satellite ETGs of the same cluster from $z=0.85$ to $z=0$, that belonged to both $z=0$ and $z=0.85$ samples and that do not change their morphology from $z=0.85$ to $z=0$. Of the 29 ETGs at $z=0.85$, only the 8 ones cited in Section \ref{sec:evoindivual} satisfy the above condition\footnote{There are 4 other $z=0.85$ satellite ETGs that remain ETGs satellites of the clusters from $z=0.85$ to $z=0$, but they are not included in the $z=0$ sample because their $M_*$ is below the adopted cut. We reanalysed the sample by adding these 4 ETGs: the conclusions of this subsection do not change.}.
These galaxies are represented in Fig. \ref{fig:s-mev1} at $z=0$ (blue filled circles) and at $z=0.85$ (red filled circles): symbols representing the same galaxy at different redshifts are connected with arrows.
The disordered orientations of the arrows suggest that these galaxies essentially do not contribute to the evolution of the $R_{\rm e} - M_*$ relation. Over the relatively narrow stellar mass range spanned by these ETGs ($10^{10.4} <  M_* / M_\odot <10^{11.2}$), their $z=0.85$ and $z=0$ $R_{\rm e} - M_*$ best fits are indistinguishable (see Table~\ref{tab:Pearson54}).
Some of these galaxies lose a significant fraction of their stellar mass from $z=0.85$ to $z=0$: the mass loss rate is $\approx 25\%$ on average and as high as $\approx 45\%$ in one case.
These stellar mass loss rates, which we verified to be independent of the adopted definition of total stellar mass (see Section~\ref{app:massdef}), are relatively high compared to what typically found in some previous theoretical works \citep[e.g.][]{2017MNRAS.464..508B,2019MNRAS.485.2287B}, but not exceptional \citep[see][]{2016MNRAS.460.1147B,2019MNRAS.485..396V}.

The aforementioned 8 galaxies are not the only ones that remain satellites of the same simulated cluster from $z=0.85$ to $z=0$: there are 10 $z=0$ satellite ETGs that at $z=0.85$ were satellite LTGs (the ones cited in Section \ref{sec:evoindivual}). Thus they are part of the $z=0$ sample, but not of the $z=0.85$ sample. We verified that the addition of these 10 galaxies has only a minor effect on the evolution of the $R_{\rm e} - M_*$ relation.

\subsubsection{ETGs that merge with the brightest cluster galaxy}
\label{sec:mergeBCG}
\begin{figure}
\includegraphics[width=\columnwidth]{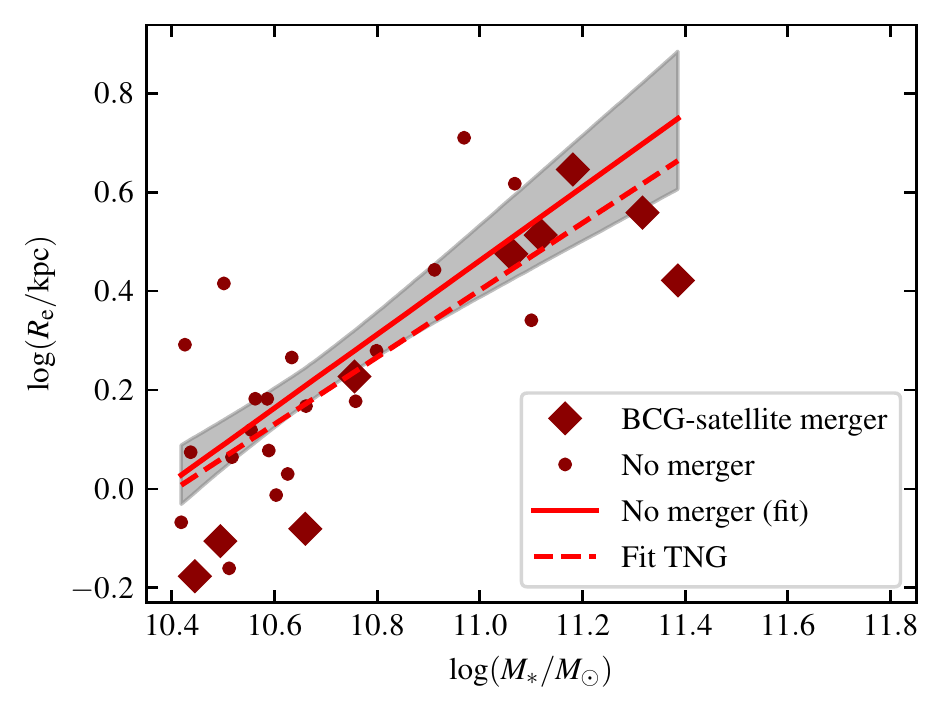}
\caption{Effective radius $R_{\rm e}$ as a function of stellar mass $M_*$ at $z=0.85$ for the 9 satellite ETGs of the simulated clusters that merge with the BCG between $z=0.85$ and $z=0$ (red diamonds), and the ones that do not (red filled circled). The median size-stellar mass relation of the red filled circles, along with its $68\%$ uncertainty (posterior highest density interval as a grey band), is a solid red line. This is compared with the median size-stellar mass relation for the entire sample of satellite ETGs of the clusters at $z=0.85$ (dashed red lines). The dashed line is the same, within the errors, as the solid one.}
\label{fig:s-mev4}
\end{figure}

Out of the 29 ETGs at $z=0.85$, 9 are cannibalized.
As shown by Fig. \ref{fig:s-mev4}, 5 of these 9 cannibalized galaxies are among the most massive satellite ETGs at $z=0.85$, consistent with the fact that, for given orbit, the dynamical friction timescale decreases for increasing mass. We note that the other 4 cannibalized galaxies, which have relatively low stellar mass ($M_*\lesssim 10^{10.8}M_\odot$ at $z=0.85$), tend to have small $R_{\rm e}$ for their stellar mass: it is probable that their compactness makes them resilient to tidal stripping and that their orbits are favourable for merging with the BCG, notwithstanding their lower mass.
We recall that throughout this work the central BCGs are never included in the sample used to study the $R_{\rm e}-M_*$ relation, so the only effect of a BCG-satellite merger is to remove satellite ETGs from our sample. To assess the effect on the $R_{\rm e}-M_*$ relation of this removal of cannibalized galaxies, we removed these 9 ETGs from the sample at $z=0.85$ and we fitted the distribution in the $M_*-R_{\rm e}$ plane of the remaining subsample of 20 ETGs at this redshift. The $z=0.85$ $R_{\rm e}-M_*$ fit of this subsample (solid red line in Fig. \ref{fig:s-mev4}) has only slightly higher normalisation than that of the entire sample (dashed red line in Fig. \ref{fig:s-mev4}), but within the errors they are the same $R_{\rm e}-M_*$ relations: this suggests that the removal of the cannibalized galaxies gives only a small contribution to the evolution of the $R_{\rm e}-M_*$ relation (see Table~\ref{tab:Pearson54}). Though not decisive for the evolution of the size-stellar mass relation, the BCG-satellite mergers have the important effect of eliminating from the cluster ETG population very dense massive galaxies that would otherwise stand out as outliers in the $R_{\rm e}-M_*$ relation at $z=0$ (see also \citealt{2019MNRAS.484..595M}).

\subsection{A more inclusive membership definition}
\label{sec:membership}
As cosmic time goes on, the Universe expands, but, within clusters of galaxies, the distances of the galaxies from the centre are expected, on average, to either decrease or remain constant \citep[e.g.][and references therein]{1980lssu.book.....P,2021MNRAS.505.5896A}.
The membership criterion implemented in TNG100 and used above becomes more inclusive with cosmic time, because the maximum distance to be a member increases as the Universe becomes less dense. In other terms, the membership criterion becomes exclusive with increasing redshift. It would be therefore useful to understand how our results are affected by adopting instead a redshift-independent criterion for membership. In this section, a high redshift galaxy is a member if it is within the physical distance considered as the maximum one of the $z=0$ sample of satellite ETGs ($\approx 4.1$ and $\approx 2.7$ Mpc for "FoF 0" and "FoF 1", respectively).
In the following we will refer to this criterion as "inclusive" and to the criterion described in Section \ref{sec:selec} as "exclusive". With the exclusive criterion we had 53 galaxies that are acquired at $z<0.85$, while 22 galaxies (i.e. $75-53$) are members at both $z=0$ and $z=0.85$. The inclusive criterion adds 19 galaxies to the 22 already counted with the exclusive criterion.
\begin{figure}
\includegraphics[width=\columnwidth]{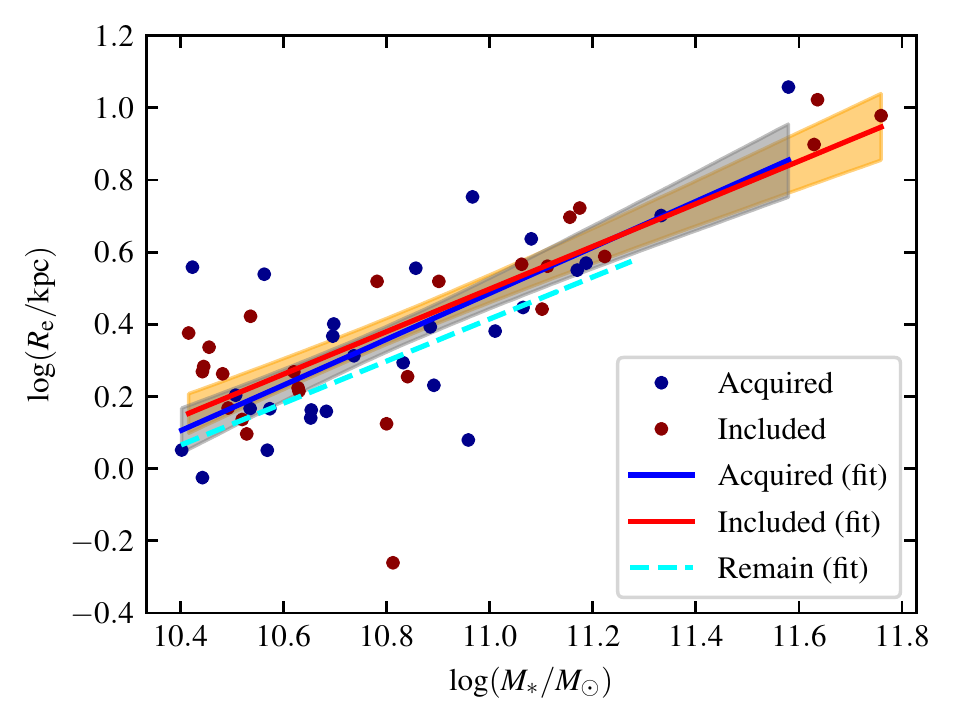}
\caption{Effective radius $R_{\rm e}$ as a function of stellar mass $M_*$ for subsamples of TNG100 galaxies at $z=0$. The blue/red circles represent the 27 galaxies that are not clusters' members according to any of the two membership criteria and 26 galaxies added according to the inclusive criterion, respectively. The two median size-stellar mass relations at $z=0$ are solid blue and red lines respectively, whereas the colored bands are their $68\%$ uncertainties (posterior highest density interval: grey for the former, orange for the latter). For comparison, the $z=0$ relation for the 22 galaxies already member at $z=0.85$ with the exclusive criterion is shown as a dashed cyan line. The blue and red solid lines are the same within the errors: they are about $1.5 \sigma$ different than the dashed line.}
\label{fig:membership}
\end{figure}

We note that at $z \approx 0.3$ the cluster "FoF 1" experiences multiple mergers. Just before the mergers its main progenitor contains $38\%$ of the present-day virial mass, with another progenitor carrying a similar amount of mass and 7 galaxies. These seven galaxies are cluster galaxies, i.e. not group or field galaxies, such as the cluster galaxies of the main progenitor of "FoF 1": there is no reason to discriminate the former from the latter. Therefore, the $z=0.85$ progenitors of these 7 galaxies (4 of which are LTGs at $z=0.85$) are also included in our high redshift comparison sample.

Figure \ref{fig:membership} shows the $R_{\rm e}-M_*$ relations at $z=0$ for the 26 galaxies added by the inclusive criterion (red circles, 9 of which are ETGs with $M_* \ge 10^{10.4} M_\odot$ at $z=0.85$) with the remaining 27 that are not clusters' members at high redshift by any of the membership criteria used so far (blue circles). The two solid lines are best fits to these data. These two fits are equal within the errors, and about 1.5$\sigma$ different from the relation for the 22 galaxies already member at $z=0.85$ with the exclusive criterion (dashed cyan line): see Table~\ref{tab:Pearson54} for details.

Therefore, there are two main drivers, both with equal importance, for the $R_{\rm e}-M_*$ relation evolution: "newcomers" (i.e. galaxies which where not members at $z=0.85$) and "suburbanites" (i.e. members living in the clusters' outskirts), both with larger size for their stellar mass. Nevertheless, only a minority ($\approx 35 \%$) of the "suburbanites" are ETGs with $M_* \ge 10^{10.4} M_\odot$ at $z=0.85$: this minority does not affect significantly the $R_{\rm e}-M_*$ relation at $z=0.85$. Our conclusion may be tested with future data looking for clustercentric dependence of the $R_{\rm e}-M_*$ relation at high redshift. Although most clusters are observed using single-pointing \emph{HST} observations (used by us in the comparison employing the exclusive membership criterion, which cover a comparable size in Mpc), some \emph{HST} mosaic should be present in the archive.

\subsection{The stellar mass function of simulated cluster ETGs}
\label{sec:StellMassFunc}

\begin{figure}
\includegraphics[width=\columnwidth]{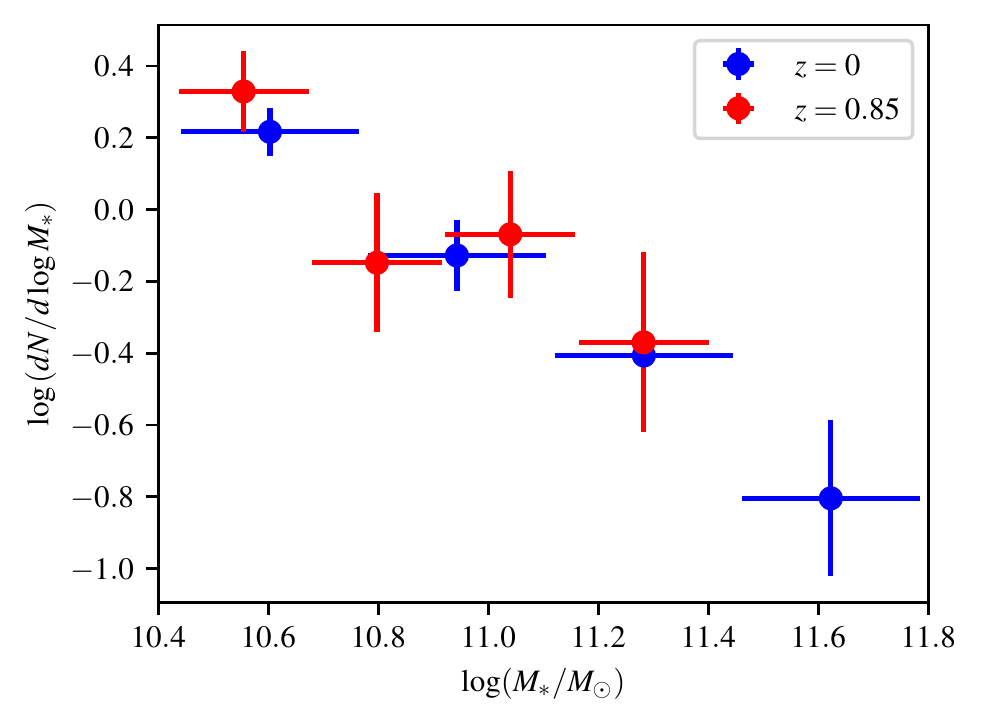}
\includegraphics[width=\columnwidth]{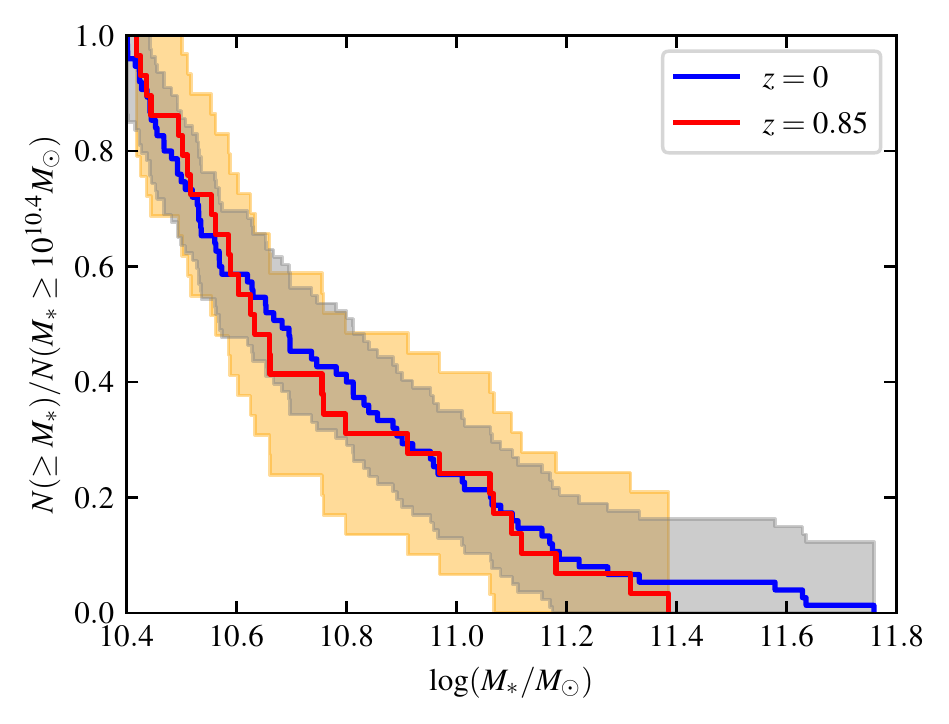}
\caption{Differential (upper panel) and cumulative (lower panel) distributions of galaxy stellar masses for the 75 ETGs at $z=0$ (blue) and the 29 ETGs at $z=0.85$ (red) belonging to our sample of simulated cluster galaxies. The distributions in both panels are normalised to the total number of ETGs at given $z$. The coloured bands in the lower panel are the $68\%$ confidence levels, generated by the Dvoretzky–Kiefer–Wolfowitz inequality \citep[][]{10.1214/aop/1176990746}, at $z=0$ (grey) and at $z=0.85$ (orange). Because of a near 1 p-value returned by the two-sample Kolmogorov-Smirnov test, the blue distribution is the same as the red one.}
\label{fig:massf}
\end{figure}

Observational studies \citep[e.g.][]{2010A&A...524A..17S,2013MNRAS.434.3469D,2014A&A...565A.120A} found that the shape of the galaxy stellar mass function of red sequence ETGs in clusters is unaltered over the past 10 Gyr, which is an additional constraint for models trying to explain the observed evolution of cluster ETGs. This behaviour of the cluster ETG population is not necessarily expected in TNG100: according to the Section \ref{sec:selec} membership criterion, most of the simulated sample at $z=0$ (53 out of 75) is composed of newly acquired galaxies, i.e.\ a subsample of objects that in principle could have belonged to populations with different stellar mass functions of cluster ETGs. In the upper panel of Fig.\ \ref{fig:massf} we plot the differential distribution of the stellar masses of simulated cluster ETGs normalised to the total number of galaxies for our $M_* \ge 10^{10.4} M_\odot$ samples at $z=0$ (blue) and $z=0.85$ (red). The corresponding cumulative distributions are shown in the lower panel of Fig.\ \ref{fig:massf}.
Given the sample size, the two stellar mass functions are consistent, because the two-sample Kolmogorov-Smirnov test returns a p-value of 0.95 (a value close to 0 would indicate differences). It follows that in these two TNG100 clusters the shape of the stellar mass function of the simulated ETGs is not significantly different at $z=0.85$ and at $z=0$. In this respect, the TNG100 simulation reproduces the observational finding. The evolution of cluster member galaxies and the acquisition of new galaxies "conspire" to maintain the shape of the galaxy stellar mass function unaltered in the explored redshift interval.

\section{Conclusions}
\label{sec:concl}
In this work we studied from a theoretical point of view the origin of the observed evolution of the $R_{\rm e}-M_*$ relation of satellite ETGs in clusters of galaxies. For this purpose we selected 75 ETGs at $ z = 0 $ with stellar mass $M_* \geq 10^{10.4} M_\odot$ belonging to the two most massive clusters of the TNG100 simulation (both with virial mass $\approx 10^{14.6} M_\odot$). We then selected 29 ETGs belonging to the main progenitors of these two clusters at $z=0.85$. We measured the $R_{\rm e}-M_*$ relations of the simulated galaxies at $z=0$ and $z=0.85$ and compared them with the relations at similar redshifts observed by \citet{2016A&A...593A...2A}. We also followed the evolution of individual simulated galaxies in the $M_*-R_{\rm e}$ plane.
Our main results can be summarized as follows:
\begin{enumerate}
\item The $R_{\rm e}-M_*$ relations found with the simulated sample are in agreement with the observations at both $z=0$ and $z=0.85$: as $z$ increases, ETGs with similar stellar mass tend to be more compact.
\item In the simulation the main drivers of the evolution of the size-stellar mass relation of satellite cluster ETGs between $z=0.85$ and $z=0$ are the acquisition by the clusters of new galaxies and the transformation of member galaxies located at $z=0.85$ at large clustercentric distances. At $z=0$ these galaxies end up being satellite ETGs on average more extended than the ETGs that already belonged to the inner parts of the clusters.
\item The shape of the stellar mass function of the simulated cluster ETGs is not significantly different at $z=0.85$ and at $z=0$, consistent with what found in observed galaxy clusters \citep[see][]{2014A&A...565A.120A}.
\end{enumerate}
Our analysis of the TNG100 simulation suggests an evolutionary scenario in which progenitor bias is dominant: according to the FoF algorithm, most of the progenitors of $z=0$ cluster ETGs were not cluster ETGs at higher $z$.
In particular, $\approx 70\%$ of current cluster satellite ETGs are acquired or long-distance galaxies and nearly half of them were LTGs at $z=0.85$: the morphological transformation of these galaxies is probably to be ascribed to interactions within the cluster, and it is thus an example of environmental quenching (e.g.\ \citealt{cimatti2019introduction}, \S 10.6). Moreover, about $30\%$ of the satellite ETGs at $z=0.85$ merge with the BCG by $z=0$ and thus they disappear from the population of satellite ETGs. This mechanism eliminates galaxies that would otherwise stand out as outliers in the $R_{\rm e}-M_*$ relation at $z=0$. We thus have three forms of progenitor bias at work: a morphological bias (galaxies that were LTGs and become ETGs), an environmental bias (galaxies that are acquired by the clusters) and a satellite/central bias (galaxies that were satellites and become centrals). Our theoretical results, in agreement with recent observational works \citep{2019MNRAS.484..595M,2020MNRAS.493.6011M}, support a scenario in which the most important effect is the environmental bias. However, we have to remark that our findings are based on only two simulated clusters of the TNG100 simulation. On the theoretical side it will be interesting to verify whether our results are confirmed when analysing other cosmological simulations and also using more clusters in order to have a better statistics at $z>0.85$ to compare simulations with higher-$z$ observational data.
On the observational side, the prediction of a dependence of the $R_{\rm e}-M_*$ relation on the clustercentric distance could be tested with wider-field observations of galaxy clusters at intermediate and high redshifts.

\section*{Acknowledgements}
We are grateful to Carlo Cannarozzo for useful discussions. We would like to thank the IllustrisTNG Project team for making their data publicly available. FM acknowledges support through the program "Rita Levi Montalcini" of the Italian MUR. MM acknowledges support from the ERC Consolidator Grant funding scheme (project ASTEROCHRONOMETRY, \hyperlink{https://www.asterochronometry.eu}{https://www.asterochronometry.eu}, G.A. n. 772293).

\section*{Data Availability}
The data underlying this article will be shared on reasonable request to the corresponding author. The simulations part of the IllustrisTNG project are publicly available at \hyperlink{https://www.tng-project.org/data/}{https://www.tng-project.org/data/}.


\bibliographystyle{mnras}
\bibliography{references} 



\appendix
\section{Evolution of the size-stellar mass relation at $\lowercase{\pmb{z > 1}}$}
\label{app:evzhigh}
In this Appendix we extend at $z>1$ the comparison between observed and simulated $R_{\rm e}-M_*$ relations of satellite cluster ETGs. The three highest redshift bins sampled by \citet{2016A&A...593A...2A} are $z_{\rm obs} = 1.32 - 1.40$, $z_{\rm obs} = 1.48 - 1.71$ and $z_{\rm obs} = 1.75 - 1.80$. For each bin, we counted the number of ETGs in the TNG100 snapshots ($z_{\rm snap}$) within its redshift range and we selected for the size-stellar mass comparison the snapshot $z_{\rm snap}$ that contains most ETGs (in case there are more than one $z_{\rm snap}$ with the same number of ETGs, we selected the snapshot closest to the average redshift of the bin). We exclude the highest-$z$ bin because the corresponding snapshot has only 4 ETGs in the clusters. The selected snapshots are at $z_{\rm snap} = 1.3$ and $z_{\rm snap} = 1.53$: their number of ETGs is 17 and 9, respectively.
\begin{figure}
\includegraphics[width=\columnwidth]{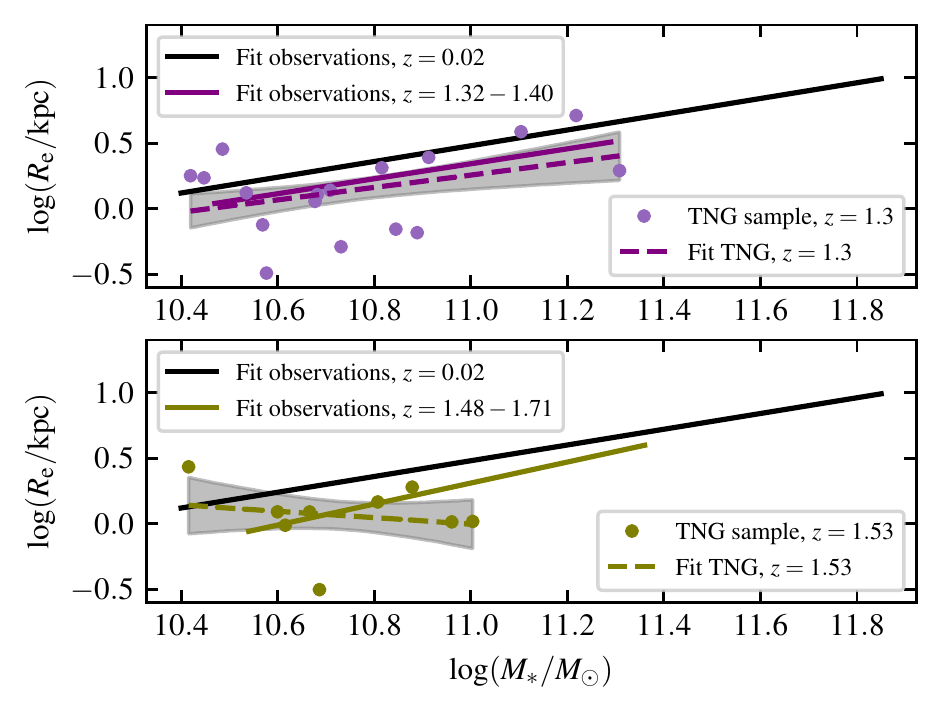}
\caption{Effective radius $R_{\rm e}$ as a function of stellar mass $M_*$ at higher redshifts than studied in the paper for our sample of TNG galaxies (filled circles). The dashed lines are the medians of the size-stellar mass relations of the distributions of the simulated ETGs: the grey bands are their $68\%$ uncertainty (posterior highest density interval). The solid lines represent \citet{2016A&A...593A...2A} best fits to their sample of observed galaxies at $z = 0.02$ (Coma cluster; black), $z = 1.32 - 1.40$ (purple) and $z = 1.48 - 1.71$ (olive). These last two translate to redshifts $z = 1.3$ (purple) and $z =1.53$ (olive) for corresponding TNG100 results. TNG100 qualitatively follows observations out to $z= 1.3$.}
\label{fig:s-mall39}
\end{figure}
For the galaxies belonging to the selected snapshots, we calculated $M_*$ and $R_{\rm e}$ as described in Section~\ref{growthcurve} and we fitted their $R_{\rm e}-M_*$ relations with the same priors as in Section \ref{sec:Sizez0}. The results are shown in Fig. \ref{fig:s-mall39}, where the dashed lines represent TNG100 median size-stellar mass relations and the solid lines the best fits of \citet{2016A&A...593A...2A}. Fig.~\ref{fig:s-mall39} shows how TNG100 qualitatively follows observations out to $z=1.3$.

\section{Total stellar mass definitions}
\label{app:massdef}
As it is well known, given that galaxies are not systems with sharp boundaries, there are different ways to define the total stellar mass of a galaxy.

In this Appendix we present the stellar mass-size plane with different total stellar mass definitions for simulated galaxies and we justify why we adopted the definition of galaxy total stellar mass $M_{*,2r_{\rm h}}$ for the $M_*-R_{\rm e}$ plane analysis of the main paper (Section \ref{sec:results}).

We consider three different definitions of the total stellar mass of a simulated galaxies: $M_{*,\rm{bound}}$, $M_{*,2r_{\rm h}}$ and $M_{*,3r_{\rm h}}$. $M_{*,\rm{bound}}$ is simply the sum of the masses of all stellar particles bound to the galaxy's subhalo.
To define the stellar mass $M_{*,2r_{\rm h}}$, we first compute the cumulative distribution $M_{\rm int}(<r)$ of stellar mass within a sphere of radius $r$. We define the 3D stellar half-mass radius $r_{\rm h}$ such that $M_{\rm int}(<r_{\rm h})=M_{*,{\rm bound}}/2$. Similarly we define the stellar mass within twice the 3D stellar half-mass radius and within three times the 3D stellar half-mass radius as $M_{*,xr_{\rm h}} \equiv M_{\rm int}(<xr_{\rm h})$ with $x=2,3$ respectively. When a definition of $M_*$ is chosen, the corresponding $R_{\rm e}$ is calculated from the growth curve so that it includes half the mass $M_*$. In Tables \ref{tab:Mhalf} - \ref{tab:Mhalf54444} we list {\sc subfind} IDs, stellar masses ($M_{*,2r_{\rm h}}$, $M_{*,\rm{bound}}$ and $M_{*,3r_{\rm h}}$) and the associated projected half-mass radii ($R_{{\rm e},2r_{\rm h}}$, $R_{{\rm e},\rm{bound}}$ and $R_{{\rm e},3r_{\rm h}}$) along the $z$-axis for our ETG sample at $z=0$ and $z=0.85$ respectively.

\begin{figure}
\includegraphics[width=\columnwidth]{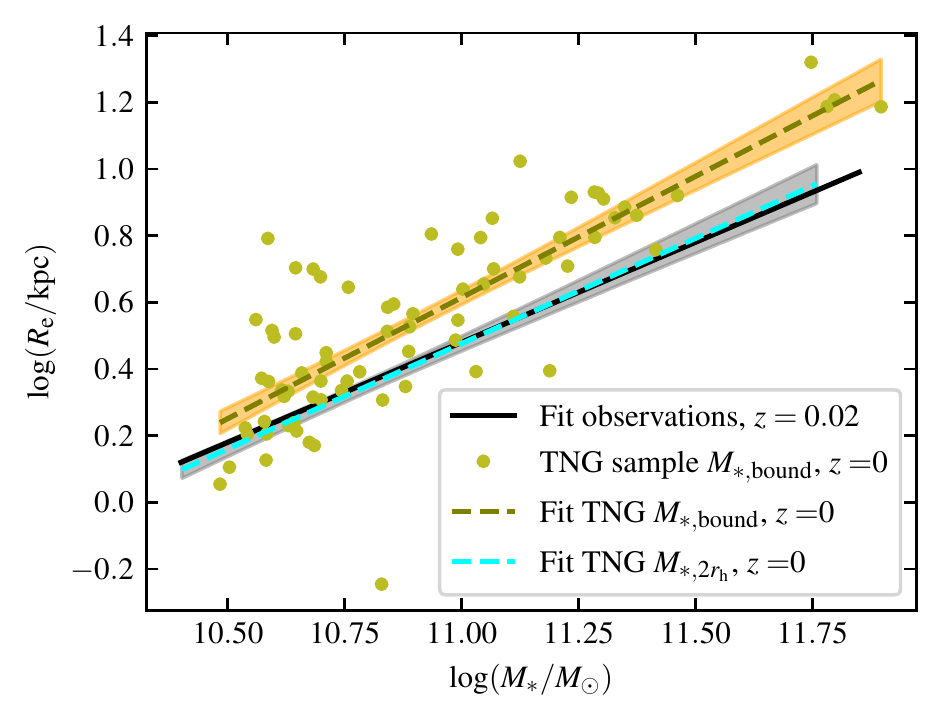}
\caption{Effective radius $R_{\rm e}=R_{\rm e,bound}$ as a function of stellar mass $M_*=M_{\rm *,bound}$ at $z=0$ for our sample of TNG ETGs (filled olive circles). Their median $R_{\rm e}-M_*$ relation (olive dashed line) along with its $68\%$ uncertainty (posterior highest density interval as an orange band) are also shown. For comparison, the $z=0$ median size-stellar mass relation for simulated data displayed in Figure \ref{fig:s-m1} is shown as a dashed cyan line along with its $68\%$ uncertainty (posterior highest density interval as a grey band). The black solid line represents \citet{2016A&A...593A...2A} best fit to the Coma cluster ETGs. The green dashed line have a higher normalisation than the black solid and cyan dashed lines.}
\label{fig:appB1}
\end{figure}

\begin{figure}
\includegraphics[width=\columnwidth]{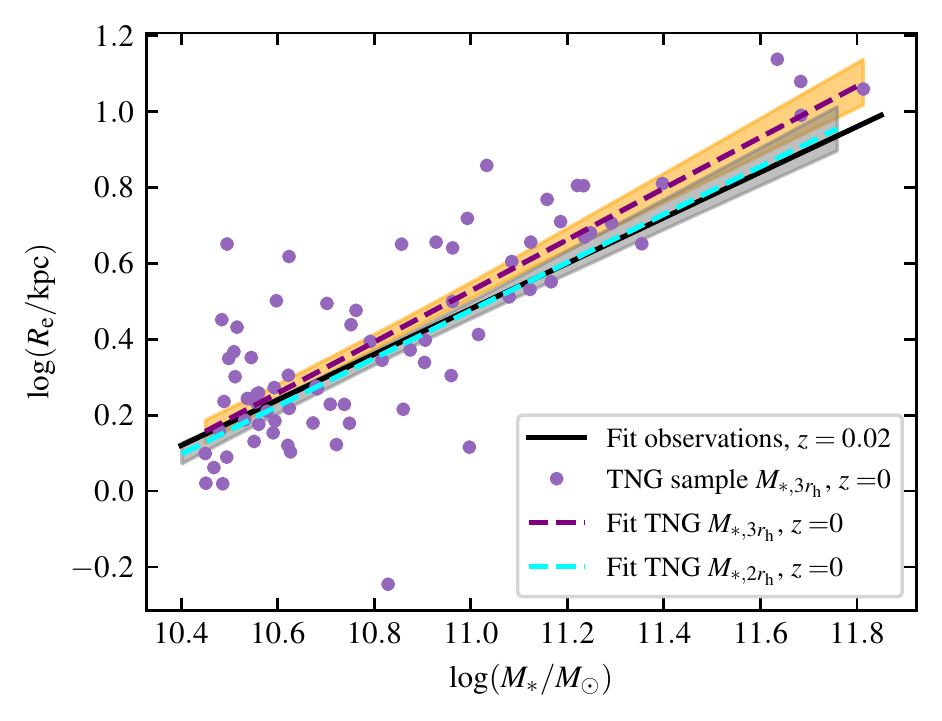}
\caption{Effective radius $R_{\rm e}=R_{{\rm e},3r_{\rm h}}$ as a function of stellar mass $M_*= M_{*,3r_{\rm h}}$ at $z=0$ for our sample of TNG ETGs (filled purple circles). Their median size-stellar mass relation (purple dashed line) along with its $68\%$ uncertainty (posterior highest density interval as an orange band) are also shown. For comparison, the $z=0$ median size-stellar mass relation for simulated data displayed in Figure \ref{fig:s-m1} is shown as a dashed cyan line along with its $68\%$ uncertainty (posterior highest density interval as a grey band). The black solid line represents \citet{2016A&A...593A...2A} best fit to the Coma cluster ETGs. The purple dashed line have a higher normalisation than the black solid and cyan dashed lines, but lower than the green dashed line of the previous figure.}
\label{fig:appB2}
\end{figure}

With $M_{*,\rm{bound}}$ and $M_{*,3r_{\rm h}}$, we analyse the stellar mass-size plane adopting the same mass selection criteria presented in Section \ref{sec:selec}, in all cases applied to $M_{*,2r_{\rm h}}$.
Figures \ref{fig:appB1} and \ref{fig:appB2} show two different median size-stellar mass relations at $z=0$ for our sample of TNG ETGs: the former adopts the stellar mass definition $M_*=M_{\rm *,bound}$ (olive dashed line), the latter adopts the stellar mass definition $M_*=M_{*,3r_{\rm h}}$ (purple dashed line). For comparison, we also show in both figures the \citet{2016A&A...593A...2A} best fit to the Coma cluster ($z=0.02$) ETGs (black solid line) and the $R_{\rm e}-M_*$ relation at $z=0$ for the TNG100 ETGs of Figure \ref{fig:s-m1}, i.e. ETGs with the stellar mass definition $M_*=M_{*,2r_{\rm h}}$. All the best fit coefficients of the size-stellar mass relations and related errors are listed in Table \ref{tab:Pearson}.
From Figures \ref{fig:s-m1}, \ref{fig:appB1} and \ref{fig:appB2} and Table \ref{tab:Pearson} follows that
\begin{enumerate}
\item assuming $M_* = M_{\rm *,bound}$ for the simulated galaxies, TNG100 galaxies tend to be more extended than real galaxies of similar stellar mass (a factor $\approx 1.3$ larger $R_{\rm e }$ for a stellar mass of $10^{10.75} M_\odot$);
\item assuming $M_*=M_{*,3r_{\rm h}}$, the size-stellar mass relation of TNG100 galaxies is more similar to the observed one, but $R_{\rm e }$ still tends to be too large for given $M_*$;
\item the relation found with the stellar mass definition $M_*=M_{*,2r_{\rm h}}$ agrees very well with observations.
\end{enumerate}
We studied the stellar mass-size plane also at $z=0.85$ with the three aforementioned stellar mass definitions finding the same trend as for the $z=0$ plane analysis. The ratio between the effective radius at given $M_*$ at $z=0$ and at $z=0.85$ is essentially independent of the definition of $M_*$: for a stellar mass of $10^{10.75} M_\odot$, this ratio is $\approx 1.23$.

The aim of this article is studying the relative evolution with redshift of the size-stellar mass relation for simulated and real galaxies. For this purpose, it is most convenient to use an operational definition of the size and stellar mass of the simulated galaxies such that at $z=0$ they behave similarly to the real ones in the $M_*-R_{\rm e}$ plane. This is why we adopt $R_{\rm e}=R_{{\rm e},2r_{\rm h}}$ and $M_*=M_{*,2r_{\rm h}}$ to compare real and simulated ETGs in our paper. We note that the same definitions of $R_{\rm e}$ and $M_*$ are used in TNG100 to broadly reproduce the observed size-stellar mass relation at $z=0$, and in other size evolution studies \citep[e.g.][]{2018MNRAS.474.3976G}.
Finally, we want to remark that the choice of $M_{*,2r_{\rm h}}$ as total stellar mass of the simulated galaxy is not necessarily the stellar mass that one would infer from a mock observation. The fact that, adopting $M_{*,2r_{\rm h}}$, we get a remarkable agreement with the observed $R_{\rm e}-M_*$ relation of ETGs at $z=0$ does not necessarily mean that TNG100 ETGs are fully realistic in terms of stellar mass and size \citep[see e.g.][]{2018MNRAS.474.3976G,2019MNRAS.484..869V}.
\begin{table*}
\centering
\caption{{\sc subfind} IDs, stellar masses and the associated projected half-mass radii of our satellite cluster ETGs sample belonging to the most massive TNG100 cluster at $z=0$ ("FoF 0"). In particular, we report the mass of all stellar particles within twice the 3D stellar half-mass radius ($M_{*,2r_{\rm h}}$), the mass of all stellar particles bound to the subhalo ($M_{*,\rm{bound}}$) and the mass of all stellar particles within three times the 3D stellar half-mass radius ($M_{*,3r_{\rm h}}$). A \citet{2003PASP..115..763C} IMF is assumed. Masses are in $M_\odot$ and effective radii in kpc.}
\label{tab:Mhalf}
\begin{tabular}{@{}lllllll@{}}
\toprule
ID & $\log M_{*,2r_{\rm h}}$ & $\log R_{{\rm e},2r_{\rm h}}$ & $\log M_{*,\rm{bound}}$ & $\log R_{{\rm e},\rm{bound}}$ & $\log M_{*,3r_{\rm h}}$ & $\log R_{{\rm e},3r_{\rm h}}$\\ \midrule
1   & 11.58 & 1.06    & 11.75 & 1.32    & 11.63 & 1.14 \\
2   & 11.63 & 0.90    & 11.80 & 1.21    & 11.68 & 0.99 \\
3   & 11.08 & 0.64    & 11.23 & 0.91    & 11.16 & 0.77 \\
4   & 10.97 & 0.75    & 11.12 & 1.02    & 11.03 & 0.86  \\
5   & 11.11 & 0.56    & 11.28 & 0.93    & 11.19 & 0.71  \\
6   & 11.33 & 0.70    & 11.46 & 0.92    & 11.40 & 0.81 \\
7   & 11.10 & 0.44    & 11.28 & 0.79    & 11.17 & 0.55  \\
8   & 11.19 & 0.57    & 11.35 & 0.89    & 11.25 & 0.68 \\
9   & 11.17 & 0.72    & 11.30 & 0.91    & 11.23 & 0.80  \\
10  & 11.22 & 0.59    & 11.37 & 0.86    & 11.29 & 0.71  \\
11  & 11.28 & 0.52    & 11.42 & 0.76    & 11.35 & 0.65 \\
13  & 10.80 & 0.12    & 10.99 & 0.49    & 10.86 & 0.22  \\
14  & 11.06 & 0.45    & 11.23 & 0.71    & 11.12 & 0.53  \\
15  & 10.70 & 0.37    & 10.85 & 0.59    & 10.75 & 0.44  \\
16  & 10.81 & 0.29    & 11.00 & 0.64    & 10.87 & 0.37  \\
17  & 10.74 & 0.31    & 10.90 & 0.57    & 10.79 & 0.39  \\
18  & 10.78 & 0.52    & 10.94 & 0.80    & 10.86 & 0.65 \\
19  & 11.01 & 0.47    & 11.12 & 0.68    & 11.08 & 0.60 \\
21  & 10.92 & 0.61    & 11.04 & 0.79    & 10.99 & 0.72 \\
22  & 10.52 & 0.14    & 10.68 & 0.32    & 10.56 & 0.18  \\
23  & 10.89 & 0.23    & 11.03 & 0.39    & 10.96 & 0.30  \\
25  & 10.67 & 0.05    & 10.83 & 0.31    & 10.72 & 0.12  \\
26  & 10.54 & 0.42    & 10.70 & 0.68    & 10.60 & 0.50  \\
27  & 10.46 & 0.34    & 10.64 & 0.70    & 10.52 & 0.43  \\
30  & 10.56 & 0.24    & 10.71 & 0.42    & 10.62 & 0.30 \\
31  & 10.75 & 0.25    & 10.89 & 0.45    & 10.82 & 0.34  \\
32  & 10.48 & 0.26    & 10.64 & 0.51    & 10.55 & 0.35 \\
33  & 10.44 & 0.27    & 10.59 & 0.52    & 10.51 & 0.37 \\
34  & 10.53 & 0.20    & 10.71 & 0.45    & 10.59 & 0.27 \\
35  & 10.47 & 0.17    & 10.62 & 0.34    & 10.54 & 0.24  \\
45  & 10.49 & 0.19    & 10.66 & 0.39    & 10.56 & 0.26  \\
53  & 10.50 & 0.07    & 10.63 & 0.23    & 10.55 & 0.13  \\
55  & 10.51 & 0.20    & 10.63 & 0.33    & 10.56 & 0.26  \\
57  & 10.43 & 0.17    & 10.59 & 0.36    & 10.49 & 0.24  \\
58  & 10.44 & 0.03    & 10.58 & 0.21    & 10.49 & 0.09  \\
60  & 10.53 & 0.12    & 10.64 & 0.24    & 10.59 & 0.18 \\
63  & 10.53 & 0.10    & 10.65 & 0.21    & 10.59 & 0.15 \\
66  & 10.42 & 0.11    & 10.54 & 0.22    & 10.48 & 0.16 \\
87  & 10.40 & -0.01    & 10.50 & 0.11    & 10.47 & 0.06 \\ 
92  & 10.40 & -0.03    & 10.48 & 0.05    & 10.45 & 0.02 \\ \bottomrule
\end{tabular}
\end{table*}
\begin{table*}
\centering
\caption{Same as Table \ref{tab:Mhalf}, but for our satellite cluster ETGs sample belonging to the second most massive TNG100 cluster at $z=0$ ("FoF 1").}
\label{tab:Mhalffff}
\begin{tabular}{@{}lllllll@{}}
\toprule
ID & $\log M_{*,2r_{\rm h}}$ & $\log R_{{\rm e},2r_{\rm h}}$ & $\log M_{*,\rm{bound}}$ & $\log R_{{\rm e},\rm{bound}}$ & $\log M_{*,3r_{\rm h}}$ & $\log R_{{\rm e},3r_{\rm h}}$\\ \midrule
17186 & 11.76 & 0.98    & 11.90 & 1.19    & 11.81 & 1.06 \\
17187 & 11.64 & 1.02    & 11.78 & 1.19    & 11.68 & 1.08 \\ 
17189 & 10.96 & 0.08    & 11.19 & 0.39    & 11.00 & 0.12 \\ 
17190 & 11.17 & 0.55    & 11.33 & 0.85    & 11.24 & 0.67 \\ 
17191 & 11.01 & 0.38    & 11.18 & 0.73    & 11.08 & 0.51\\ 
17192 & 11.16 & 0.70    & 11.29 & 0.93    & 11.22 & 0.80\\
17193 & 11.06 & 0.57    & 11.21 & 0.79    & 11.12 & 0.66\\
17195 & 10.90 & 0.52    & 11.07 & 0.85    & 10.96 & 0.64\\ 
17196 & 10.88 & 0.39    & 11.07 & 0.70    & 10.96 & 0.50\\
17197 & 10.57 & 0.17    & 10.70 & 0.31    & 10.62 & 0.22\\
17198 & 10.84 & 0.25    & 11.05 & 0.65    & 10.90 & 0.34\\
17199 & 10.83 & 0.29    & 10.99 & 0.55    & 10.91 & 0.40\\
17200 & 10.70 & 0.40    & 10.84 & 0.58    & 10.76 & 0.48\\
17201 & 10.57 & 0.05    & 10.74 & 0.34    & 10.62 & 0.12\\
17202 & 10.95 & 0.33    & 11.11 & 0.56    & 11.02 & 0.41\\
17203 & 10.65 & 0.14    & 10.84 & 0.51    & 10.71 & 0.23\\
17204 & 10.86 & 0.56    & 10.99 & 0.76    & 10.93 & 0.66\\
17206 & 10.70 & 0.13    & 10.88 & 0.35    & 10.75 & 0.18\\
17207 & 10.63 & 0.22    & 10.75 & 0.36    & 10.68 & 0.28\\
17208 & 10.68 & 0.16    & 10.89 & 0.53    & 10.74 & 0.23\\
17209 & 10.40 & 0.05    & 10.54 & 0.20    & 10.45 & 0.10\\
17211 & 10.56 & 0.54    & 10.68 & 0.70    & 10.62 & 0.62\\
17212 & 10.62 & 0.27    & 10.76 & 0.65    & 10.70 & 0.49\\
17213 & 10.63 & 0.21    & 10.78 & 0.39    & 10.68 & 0.27\\
17216 & 10.42 & 0.56    & 10.59 & 0.79    & 10.50 & 0.65 \\
17218 & 10.45 & 0.28    & 10.60 & 0.50    & 10.50 & 0.35\\
17219 & 10.54 & 0.17    & 10.70 & 0.36    & 10.58 & 0.21\\
17220 & 10.81 & -0.26    & 10.83 & -0.25  & 10.83 & -0.25\\
17222 & 10.42 & 0.38    & 10.56 & 0.55    & 10.48 & 0.45\\
17223 & 10.49 & 0.17    & 10.62 & 0.32    & 10.55 & 0.23\\
17224 & 10.57 & 0.04    & 10.68 & 0.17    & 10.63 & 0.10\\
17225 & 10.44 & -0.02    & 10.58 & 0.13   & 10.49 & 0.02 \\
17227 & 10.65 & 0.16    & 10.67 & 0.18    & 10.67 & 0.18\\
17228 & 10.45 & 0.24    & 10.57 & 0.37    & 10.51 & 0.30\\
17230 & 10.47 & 0.12    & 10.58 & 0.24    & 10.53 & 0.19\\\bottomrule
\end{tabular}
\end{table*}
\begin{table*}
\centering
\caption{Same as Table \ref{tab:Mhalf} and Table \ref{tab:Mhalffff}, but for our $z=0.85$ sample of satellite cluster ETGs.}
\label{tab:Mhalf54444}
\begin{tabular}{@{}lllllll@{}}
\toprule
ID & $\log M_{*,2r_{\rm h}}$ & $\log R_{{\rm e},2r_{\rm h}}$ & $\log M_{*,\rm{bound}}$ & $\log R_{{\rm e},\rm{bound}}$ & $\log M_{*,3r_{\rm h}}$ & $\log R_{{\rm e},3r_{\rm h}}$\\ \midrule
1  & 11.18 & 0.65    & 11.34 & 0.97  & 11.22 & 0.71 \\
3  & 11.12 & 0.51    & 11.28 & 0.84  & 11.17 & 0.61  \\
4  & 11.32 & 0.56    & 11.43 & 0.75  & 11.40 & 0.69  \\
7  & 10.97 & 0.71    & 11.08 & 0.88  & 11.03 & 0.81  \\
8  & 10.60 & -0.01    & 10.81 & 0.26  & 10.64 & 0.04  \\
9  & 11.07 & 0.62    & 11.17 & 0.77  & 11.13 & 0.72  \\
10 & 10.51 & -0.16    & 10.68 & 0.10  & 10.56 & -0.10 \\
11 & 10.91 & 0.44    & 11.07 & 0.72  & 10.98 & 0.55  \\
13 & 11.06 & 0.48    & 11.16 & 0.62  & 11.14 & 0.60 \\
14 & 10.80 & 0.28    & 10.98 & 0.62  & 10.85 & 0.36 \\
15 & 10.59 & 0.18    & 10.77 & 0.45  & 10.65 & 0.26  \\
17 & 10.76 & 0.23    & 10.93 & 0.53  & 10.83 & 0.34  \\
20 & 10.44 & 0.07    & 10.60 & 0.29  & 10.49 & 0.13 \\
21 & 10.66 & 0.17    & 10.84 & 0.44  & 10.73 & 0.25 \\ 
23 & 10.56 & 0.18    & 10.74 & 0.42  & 10.63 & 0.26 \\ 
27 & 10.66 & -0.08    & 10.82 & 0.14  & 10.73 & 0.01 \\
28 & 10.63 & 0.27    & 10.77 & 0.45  & 10.70 & 0.35 \\
29 & 10.52 & 0.06    & 10.69 & 0.35  & 10.58 & 0.15 \\
36 & 10.55 & 0.12    & 10.68 & 0.25  & 10.61 & 0.18 \\
39 & 10.43 & 0.29    & 10.54 & 0.45  & 10.49 & 0.37 \\
42 & 10.42 & -0.07    & 10.57 & 0.14  & 10.49 & 0.01 \\
47 & 10.49 & -0.11    & 10.58 & -0.03 & 10.56 & -0.04 \\
50 & 10.44 & -0.18    & 10.53 & -0.11 & 10.50 & -0.13 \\
27739 & 11.39 & 0.42    & 11.56 & 0.77  & 11.46 & 0.54 \\
27741 & 11.10 & 0.34    & 11.27 & 0.67  & 11.17 & 0.46 \\
27743 & 10.76 & 0.18    & 10.91 & 0.44  & 10.83 & 0.29 \\
27744 & 10.63 & 0.03    & 10.77 & 0.23  & 10.68 & 0.10 \\
27746 & 10.59 & 0.08    & 10.73 & 0.27  & 10.65 & 0.15 \\
27748 & 10.50 & 0.42    & 10.62 & 0.56  & 10.57 & 0.49 \\ \bottomrule
\end{tabular}
\end{table*}
\begin{table*}
\centering
\caption{The mean slope $\alpha$, intercept $\gamma$ and intrinsic scatter $\sigma$ of the linear relation $\log (R_{\rm e}/\mbox{kpc})=\alpha \left[ \log (M_*/M_\odot) - 10.75 \right]+\gamma$, along with their uncertainties, of the $z=0$ best-fitting relations of simulated galaxies displayed in Figures \ref{fig:appB1} and \ref{fig:appB2}.} 
\label{tab:Pearson}
\begin{tabular}{@{}llll@{}}
\toprule
Relation at $z=0$                                  & $\alpha$          & $\gamma$ & $\sigma$\\ \midrule
log$R_{{\rm e},2r_{\rm h}}$-log$M_{*,2r_{\rm h}}$  & $0.63 \pm 0.06$   & $0.318 \pm 0.019$     &  $0.161 \pm 0.014$\\ 
log$R_{\rm e,bound}$-log$M_{\rm *,bound}$          & $0.73 \pm 0.06$   & $0.43 \pm 0.02$       & $0.177 \pm 0.015$\\
log$R_{{\rm e},3r_{\rm h}}$-log$M_{*,3r_{\rm h}}$  & $0.67 \pm 0.06$   & $0.36 \pm 0.02$       & $0.169 \pm 0.014$\\
Coma cluster ($z=0.02$)                            & $0.60 \pm 0.06$   & $0.33 \pm 0.02$       &  $0.16 \pm 0.01$\\ \bottomrule
\end{tabular}
\end{table*}


\bsp	
\label{lastpage}
\end{document}